\documentclass[twocolumn,aps,prb]{revtex4}
\usepackage{amsmath,amssymb,mathrsfs,bm}
\usepackage{xcolor}
\usepackage{graphicx,dcolumn,times}
\begin{document}
\title{Approaching the standard quantum limit of a Rydberg-atom microwave electrometer}
\author{Hai-Tao Tu$^{1}$\footnotemark[2], Kai-Yu Liao$^{1,2}$\footnote[2]{These authors contributed equally to this work.}\footnotemark[1], Guo-Dong He$^{1}$, Yi-Fei Zhu$^{1}$, Si-Yuan Qiu$^{1}$, Hao Jiang$^{1}$, Wei Huang$^{1}$, Wu Bian$^{1}$, Hui Yan$^{1,2,3,4}$\footnotemark[1] and Shi-Liang Zhu$^{1,2,3}$\footnote[1]
{email: kaiyu.liao@m.scnu.edu.cn;yanhui@scnu.edu.cn;slzhu@scnu.edu.cn}}

\affiliation {$^1$ Key Laboratory of Atomic and Subatomic Structure and Quantum Control (Ministry of Education), School of Physics, South China Normal University, Guangzhou 510006, China}

\affiliation {$^2$ Guangdong Provincial Key Laboratory of Quantum Engineering and Quantum Materials, South China Normal University, Guangzhou 510006, China}

\affiliation {$^3$ Guangdong-Hong Kong Joint Laboratory of Quantum Matter, Frontier Research Institute for Physics, South China Normal University, Guangzhou 510006, China}

\affiliation{ $^4$ GPETR Center for Quantum Precision Measurement, South China Normal University, Guangzhou 510006, China}

\begin{abstract}
\textbf{The development of a microwave electrometer with inherent uncertainty approaching its ultimate limit carries both fundamental and technological significance\cite{DegenRMP2017, QmRMP2018}. Recently, the Rydberg electrometer has garnered considerable attention  due to its exceptional sensitivity\cite{JingNP2020, PrajapatiAPL2021}, small-size\cite{MicroNP2010, MarylandPRL2018}, and broad tunability\cite{MeyerJPB2019}. This specific quantum sensor utilizes low-entropy laser beams to detect disturbances in atomic internal states, thereby circumventing the intrinsic thermal noise encountered by its classical counterparts \cite{IEEEreview2021, CompNoiseT2022}. However, due to the thermal motion of atoms\cite{MeyerPRA2021}, the advanced Rydberg-atom microwave electrometer falls considerably short of the standard quantum limit by over three orders of magnitude. In this study, we utilize an optically thin medium with approximately 5.2$\times$10$^{5}$ laser-cooled atoms\cite{LiaoPRA2020} to implement heterodyne detection.  By mitigating a variety of noises and strategically optimizing the parameters of the Rydberg electrometer, our study achieves an electric-field sensitivity of 10.0 nV cm$^{-1}$ Hz$^{-1/2}$ at a 100 Hz repetition rate, reaching a factor  of 2.6 above the standard quantum limit and  a minimum detectable field of 540 pV cm$^{-1}$.  We also provide an in-depth analysis of noise mechanisms and determine optimal parameters to bolster the performance of Rydberg-atom sensors. Our work provides insights into the inherent capacities and limitations of Rydberg electrometers, while offering superior sensitivity for detecting weak microwave signals in numerous applications.}
\end{abstract}
\maketitle


The optimal precision in quantum metrology using non-entangled states is inversely proportional to the square root of the quantity $N$ of quantum particles involved in the measurement. This relation is known as the standard quantum limit (SQL)\cite{BecerraSQL2013, IEEESQL2022}. In recent years, by employing a bright atomic resonance within the electromagnetically induced transparency (EIT) signal, Rydberg atoms have provided a self-calibrating\cite{ShafferNP2012, HollowayIEEE2014} and ultra-sensitive platform for microwave (MW) electric field measurements. According to a semiclassical model\cite{ShafferJPB2015}, Rydberg sensors detect an electric field $E$ by observing a frequency shift $\delta$ on the Rydberg state, provided by $\delta=\mu_{\mathrm{MW}}E/h$, where $h$ is the Planck's constant, and $\mu_{\mathrm{MW}}$ is the dipole moment of Rydberg transition. The phase  $\phi = \delta T_m$ evolves during a measurement time $T_m$. When decoherence is present, the single-shot measurement cannot exceed the dephasing time $T_{2}$ of the EIT process\cite{IEEEreview2021}, and the resolvable  phase shift is constrained by atom shot noise, resulting in $\Delta \phi_{\mathrm{min}}=1 / \sqrt{N T'/T_{2}}$ over an overall integration period $T'$. Consequently, the atom-shot-noise limited sensitivity, or the SQL of a Rydberg electrometer, takes the form,
\begin{eqnarray}
\begin{array}{ll}
\frac{E_{\mathrm{SQL}}}{\sqrt{\mathrm{Hz}}}=\frac{h}{\mu_{\mathrm{MW}} \sqrt{N T_2}},
\end{array}
\end{eqnarray}
by setting the signal-to-noise ratio $\phi/\Delta \phi_{\mathrm{min}} = 1$ and $T'$ = 1s.

The sensitivity of the Rydberg electrometer is primarily degraded by the thermal motion of atomic gas, from three primary aspects. Firstly, atoms at a finite temperature undergo Doppler shifts that detune the control lasers from resonance. This phenomenon results in an averaging effect that reduces the transmission of the EIT spectrum, thereby decreasing the photon-shot-noise (PSN)-limited sensitivity\cite{MeyerPRA2021}. Secondly, the departure of thermal atoms from the laser beams and  subsequently compensated by the introduction of new ground-state atoms from outside of the laser beams  will cause a loss in the Rydberg population and the coherence between states, resulting in transit dephasing that degrades the atom-shot-noise-limited sensitivity. Most notably, the transit of the thermally excited atoms across the laser beams introduces a random disturbance of optical field scattering through the atomic sample. Particularly within the frequency range of 1-100 kHz, the atomic transit noise demonstrates itself as the major source of noise in the optical readout stage\cite{JingNP2020}, compared to other noise sources such as photodetection and laser intensity fluctuation.

These limitations mean that prior work on the Rydberg-atom electrometer, which employs a thermal vapour sample of atoms, has not yet achieved optimal performance\cite{ShanxiOE2023}. Notably, the state-of-art sensitivities currently available are more than three orders of magnitude inferior to their respective SQLs. Furthermore, the existence of a significant deviation between the SQL and the observed sensitivity suggests that the exact noise mechanism in Rydberg sensors has not been confirmed. As such, the determination of optimal parameters to boost the current performance remains an unresolved issue.

\begin{figure*}[tp]
\begin{center}
\includegraphics[width=17cm]{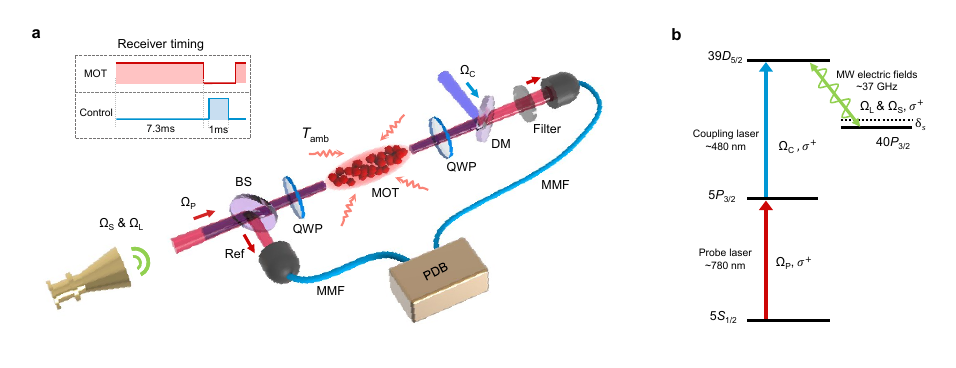}
\caption{\label{fig:setup} \textbf{Schematic representation of a cold Rydberg-atom receiver. }
\textbf{a}, Experimental setup. Three circularly-polarized control fields (780 nm: \emph{probe} laser, 481 nm: \emph{coupling} laser and 36.9 GHz: \emph{local} MW) are applied to a cold $^{87}$Rb atomic ensemble at moderate Rabi frequencies, serving as a Rydberg-atom mixer for the detection of the weak \emph{signal} MW. The counterpropagating probe and coupling beams are combined with a dichroic mirror (DM), and then concentrated at the center of a 2D magneto-optical trap (MOT). After a beam splitter (BS), probe laser beam is separated from a reference (Ref) beam, and subsequently two beams are coupled into a multi-mode fiber (MMF) for optical balanced detection. The top left shows the operation timing comprising the MOT sequence and the control laser profile in each experimental cycle. The Rydberg-atom MW receiver operating in a free space configuration is affected concurrently by internal and
external noise sources. These include atomic shot noise, PSN of the probe field, photodetector noise, fluctuations of the vacuum MW field and black-body radiation (light red wavy) at an ambient temperature of $T_{\mathrm{amb}}$ = 293 K. \textbf{b}, $^{87}$Rb energy level structure employed in MW heterodyne detection. }
\end{center}
\end{figure*}

At present, quantum sensors based  on cold atoms deliver cutting-edge performance in a wide range of precision measurement applications, these include gravimeters\cite{GravityN2014}, clocks\cite{ClockRMP2015}, and inertial sensors\cite{DuttaPRL2016, GeigerQS2020}. This proficiency is facilitated by the minimized velocity distribution of cold atoms, which allows for extended interrogation times. By applying laser cooling to eliminate atomic transit noise, we can closely approach  the SQL of Rydberg sensors.
In this study, we introduce the inaugural experimental display of a heterodyne receiver armed with laser-cooled atoms and shed light on the noise mechanisms of the Rydberg receiver. Operating under ambient conditions at room temperature, the Rydberg receiver via a thin optical sample exhibits superior field sensitivity of 10.0 nV cm$^{-1}$ Hz$^{-1/2}$. This sensitivity enhancement indicates a threefold improvement over the preceding atomic heterodyne receiver based on vapor cell, and is merely a factor of 2.6 above its SQL. Furthermore, this quantum receiver attains a minimum detectable MW field down to 540 pV cm$^{-1}$, within a temporal measurement span of 420 s and an instantaneous bandwidth of 2.3 MHz. We present conclusive evidence underlining a specific methodology for sensitive operations, indicating that, upon reducing the atom heating effect, a cold Rydberg receiver can outstrip the thermal noise threshold of traditional receivers\cite{IEEEreview2021} with a thicker medium.

\bigskip
\noindent\textbf{Results}\\
\noindent\textbf{Noise mechanism.} We now theoretically describe the noise mechanism that restricts the sensitivity of Rydberg receivers. Detailed information regarding this can be found in the methods section and supplementary section 1. Given that the minimal detectable electric field scales with the square root of observation bandwidth $\Delta f$, the \emph{noise equivalent field} (NEF) is defined by the ratio\cite{CompNoiseT2022}
  \begin{eqnarray} \mathrm{NEF}=\frac{\left|E_{\mathrm{MW}}\right|_{\min }}{\sqrt{\Delta f}}. \end{eqnarray}
This ratio effectively quantifies the achievable sensitivity pertaining to electric-field sensing.

A Rydberg-atom MW sensor is influenced by both internal and external sources of noise. The internal sources include the probe fluctuations induced by atomic shot noise, labeled as $\mathrm{{NEF}_{at}}=\frac{E_{\mathrm{SQL}}}{\sqrt{\mathrm{Hz}}}$, and practical noise sources, like PSN in the probe beam $\mathrm{{NEF}_{ph}}$, as well as photodetector noise $\mathrm{{NEF}_{pd}}$.
Typically, atomic sensors operates with sufficient power in the detected optical field, meaning the electronics' noise floor does not serve as the primary noise source in the optical detection path.  Additionally, the shot noise sources associated with the atoms and optical fields are typically coupled: an increased power probe laser leads to high population of atoms in the Rydberg state, while Rydberg-Rydberg interactions decrease the coherence time $T_{2}$ for electric-field sensing. The internal noise floor of atomic sensor can be expressed as
\begin{small}
\begin{eqnarray}
\begin{array}{ll}
\mathrm{{NEF}_{in}}=\sqrt{\mathrm{NEF}_{\mathrm{at}}^2+2r\mathrm{NEF}_{\mathrm{at}}\mathrm{NEF}_{\mathrm{ph}}+\mathrm{NEF}_{\mathrm{ph}}^2+\mathrm{NEF}_{\mathrm{pd}}^2},
\end{array}
\end{eqnarray}
\end{small}
where $r$ is the correlation coefficient between the noise equivalent fields $\mathrm{{NEF}_{at}}$ and $\mathrm{{NEF}_{ph}}$.

\begin{figure*}[tb]
\begin{center}
\includegraphics[width=17cm]{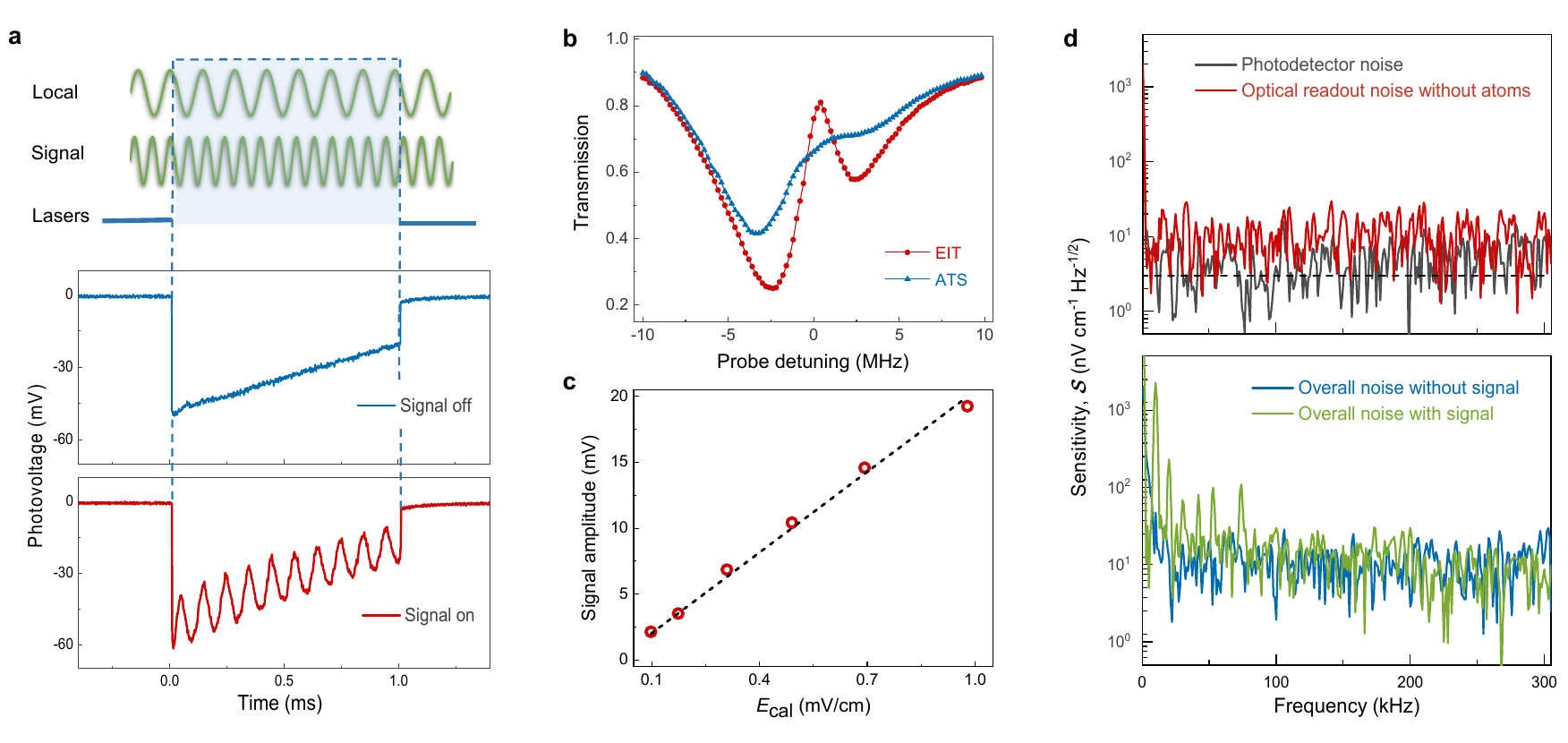}
\caption{ \textbf{Proof-of-principle MW heterodyne detection on cold atoms.} \textbf{a}, Time traces of the PDB photovoltage with continuous microwaves and square-pulsed control lasers. When the signal MW is turned off (upper), a slowly rising DC signal is evident in the probe laser transmission, indicating the decreasing optical depth of the atomic cloud during the detection period. With the signal MW turned on (lower), the time trace similarly displays an additional oscillation, the peak-to-peak amplitude of which is double that of the heterodyne sensing signal. \textbf{b}, Normalized transmission as a function of probe laser detuning. Both the EIT (red circle) and AT-splitting (blue triangle) spectra are recorded at the optimal point of the experiment. \textbf{c}, Heterodyne signal amplitude as a function of the MW electric field $E_\mathrm{cal}$. The electric field value, $E_\mathrm{cal}$, is calibrated using the standard antenna method, and the dotted line indicates the linear fit. \textbf{d}, Sensitivity spectra of the cold Rydberg-atom receiver. A sensitivity spectrum, normalized according to the 1 kHz resolution bandwidth, is measured for  the photodetector noise (dark grey curve), the optical readout noise (red curve), and the overall noise of atomic heterodyne when a signal MW at 10 kHz intermediate frequency is turned on (green curve) or turned off (blue curve). The red curve arises from the measurement made by removing the atomic cloud and maintaining the probe laser beam in the same condition as the overall noise measurement. The black dashed line denotes the photodetector's noise floor level, which assists in calibrating the spectral density of the photovoltage during the DFT process. }
\end{center}
\end{figure*}

In the context of a free-space coupled atomic receiver, the external noise sources encompass black-body radiation from the surrounding environment at temperature $T_{\mathrm{amb}}$, as well as vacuum fluctuations of the MW field. The atomic heterodyne, as a coherent detection methodology, retrieves the quadratures of the thermally-generated electromagnetic field\cite{CompNoiseT2022}. According to the Callen-Welton law, the sum of thermal photons at a frequency $\nu$ and vacuum noise collectively contribute to a spectral density of electric-field amplitude. This can be written as
\begin{eqnarray}
\label{eq:BBR} \begin{array}{ll} \mathrm{NEF}_{\mathrm{ex}}=\sqrt{\frac{8 \pi h \nu^3}{ \varepsilon_0 c^3 G} [2 n_{\mathrm{th}}(\nu, T_{\mathrm{amb}})+1]}, \end{array} \end{eqnarray}
where $G$ symbolizes the effective gain of the atomic sensor, $c$ stands for the speed of light, $\varepsilon_0$ signifies the dielectric constant, and $n_{\mathrm{th}}(\nu, T_{\mathrm{amb}}) =\frac{1}{e^{h \nu / k_B T_{\mathrm{amb}}}-1}$ represents the Bose-Einstein distribution of thermal photons. Since the internal and external noise sources are independent, the sensitivity limit of a free-space coupled atomic heterodyne with relation to the overall noise is expressed as
\begin{eqnarray}
\label{eq:NEF} \begin{array}{ll} \mathcal{S}=\sqrt{\mathrm{NEF}_{\mathrm{ex}}^2+\mathrm{NEF}_{\mathrm{in}}^2}. \end{array}
\end{eqnarray}
Based on theoretical analysis and simulation of PSN sensitivity, an atomic heterodyne receiver can approach the SQL only when an optically thin sample of cold atoms is used (see Supplementary Section 2).

\bigskip
\noindent\textbf{Experimental setup.} The implementation of MW heterodyne detection experimentally utilizes a cloud of $^{87}$Rb atoms, which are prepared in a two-dimensional (2D) magneto-optical trap (MOT) as depicted in Fig. 1a. The longitudinal length of the cigar-shaped cloud is approximately 2 cm, and the atomic temperature is around 200 $\mu$K. The entire experiment operates periodically with a repetition rate of 100 Hz, incorporating a MOT loading time of 7.3 ms and a heterodyne detection window of 2.7 ms (Fig. 1a, upper left). During the detection window, laser-cooled atoms are initially prepared in the hyperfine state $|5S_{1/2}, F=2\rangle$. The quadrupole-gradient magnetic field is turned off, while the stray magnetic field is compensated by three pairs of Helmholtz coils and reduced to less than 5 mG. The relevant atomic energy level diagram for MW heterodyne detection is shown in Fig. 1b. After traversing a quarter-wave plate (QWP), the circularly polarized probe and coupling beams are concentrated on the center of the atomic cloud with $1/e^2$ radii of $500 \mu$m and $400 \mu$m, respectively. The peak Rabi frequency of the coupling beam is $\Omega^{(0)}_{\mathrm{C}}/2\pi$ = 6.1 MHz. The frequencies of probe and coupling lasers are locked to a high-finesse Fabry-P'{e}rot cavity and resonate with the transitions of $|5S{1/2}, F=2\rangle\leftrightarrow|5P_{3/2}, F=3\rangle$ and $|5P_{3/2}, F=3\rangle\leftrightarrow|39D_{5/2}\rangle$, respectively.

\begin{figure*}[tb]
\begin{center}
\includegraphics[width=18cm]{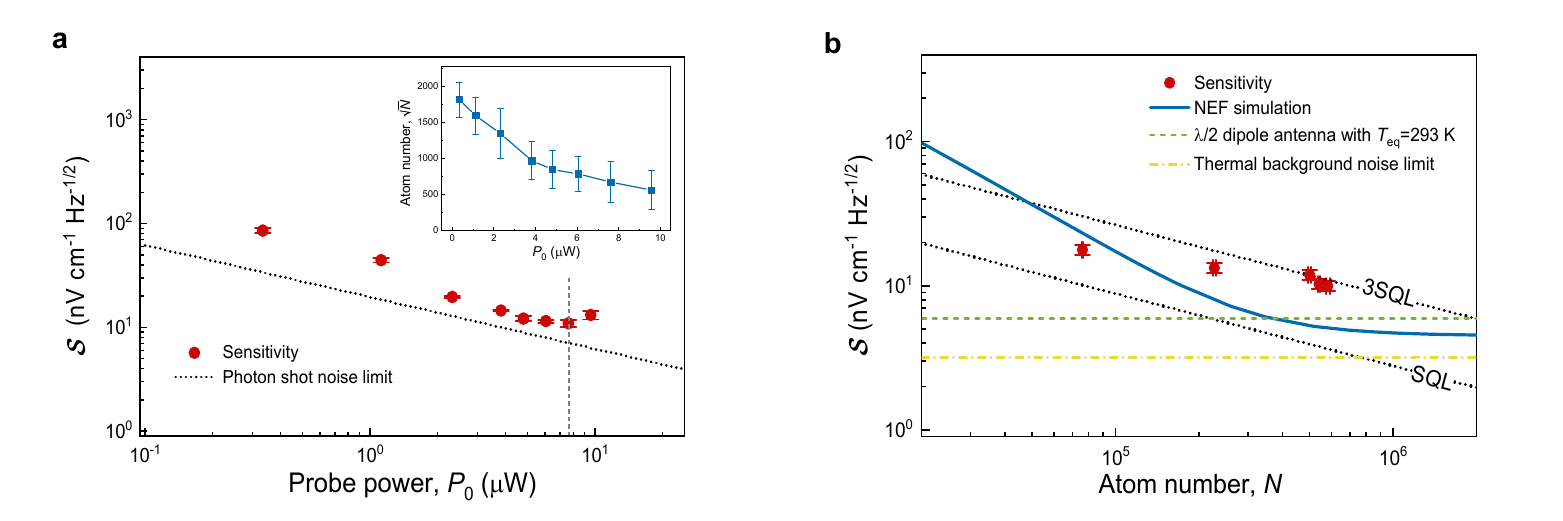}
\caption{\label{fig:SQL}  \textbf{Shot-noise limited sensitivity.} \textbf{a}, Sensitivity $\mathcal{S}$ in relation to the incident power $P$ of the probe beam. The black dotted line, deduced from Eq. (10), represents the PSN limited sensitivity for our setup. The inset illustrates the dependence of the measured atom number on the power of the probe laser. \textbf{b}, Sensitivity $\mathcal{S}$ against the number of atoms. Measurements of sensitivities are conducted at a probe power of 7.6 $\mu$W, as indicated by the vertical dashed lines in \textbf{a}. The blue curve represents the theoretical prediction according to Eq. (5). The two dotted lines symbolize the single and triple SQL sensitivities, respectively, acquired from Eq. (1) whereby the atomic coherence time is determined by Eq. (14). The dashed green line exhibits the $T_{\mathrm{eq}}$ = 293 K sensitivity constraint $\mathcal{S}_{th}$ of an optimally functioning half-wave dipole antenna within the same frequency band. The dash-dotted yellow line signifies the sensitivity bound by the thermal background ($T_{\mathrm{amb}}$ = 293 K), correlating to $\mathrm{{NEF}_{ex}}$ = 3.2 nV cm$^{-1}$ Hz$^{-1/2}$. In \textbf{a} and \textbf{b}, the measured sensitivity (red circles) is the averaged spectral density within the frequency range of 10 kHz to 300 kHz, and the error bars mark the standard deviation gathered from three measurements. The atom count is derived from 60 individual measurements of the optical depth, made through fittings to the two-level absorption profile and using the optimal probe Rabi frequency as an input in the two-level master equation.}
\end{center}
\end{figure*}

The local and signal microwaves are generated from different sources and are connected to a 10 MHz rubidium frequency reference. For a proof-of-principle demonstration, the two microwaves are combined via a resistive power divider and are subsequently coupled to the free space through a helical antenna\cite{TuNP2022}. The axial direction of the antenna is rendered collinear with the laser beams, and the MOT chamber is situated within the far-field regime. The local microwave, with a frequency around 36.9 GHz, resonates with the Rydberg transition\cite{SibalicCPC2017} of $|39D_{5/2}\rangle\leftrightarrow|40P_{3/2}\rangle$, yielding $\Omega_{\mathrm{L}}/2\pi$ = 2.0 MHz. The probe and reference beams are directed into the multi-mode fiber and detected by a balanced photodetector (PDB, Thorlabs PDB210A). A narrow bandpass filter is employed ahead of the PDB to further isolate scattering from the coupling beam. The PDB photovoltage is recorded by a digital acquisition oscilloscope (R\&S RTE1024), and the amplitude spectral density for MW sensitivity is analysed using a desktop computer. For measurements within the 3-dB bandwidth, an avalanche photodetector (Thorlabs APD130A) with a 50-MHz bandwidth replaces the photodetector.

\bigskip
\noindent\textbf{Heterodyne detection of MW field with cold atoms.}
During the detection window, we apply square-pulsed control lasers with a duration of 1 ms for atomic heterodyning. To minimize the impact of PSN, the probe laser beam maintains a moderate power level (approximately 5 $\mu$W). Despite the counterpropagation of the probe beam and coupling beam, their residual momentum leads to atom loss in the MOT. Due to both the heating effect\cite{HeatPRA1997} and the free expansion of the atomic ensemble\cite{WangNP2019}, the optical depth (OD) of the MOT for the transition $|5S_{1/2}, F=2\rangle\leftrightarrow|5P_{3/2}, F=3\rangle$ linearly decreases from 0.46 to 0.22 throughout the 1 ms heterodyne time. As depicted in Fig. 2a, the probe transmission exhibits a slow growth once initiated. This phenomenon, attributable to a varying OD, is also observed in the asymmetry of both EIT and Autler-Townes (AT) splitting spectra measured in our experiment (Fig. 2b). With the signal MW turned on, the time trace of photovoltage follows a similar rising pattern, but also displays an extra oscillation.

We next conduct a discrete Fourier transform (DFT) on the photovoltage to extract corresponding frequency and amplitude data. The extracted frequency represents the intermediate frequency of the cold Rydberg-atom receiver, whilst the extracted amplitude signifies the signal of the heterodyne detection. In order to correct the amplitude of the windowed time-signal during the DFT, we utilize the calibration information from the balanced photodetector to determine the noise equivalent voltage (NEV). Once the amplitude correction factor has been established, we can easily determine the photovoltage spectral density. As illustrated in Fig. 2c, the oscillation amplitude exhibits a linear dependence on the applied MW electric field $E_\mathrm{cal}$, resulting in a field-to-voltage sensor\cite{WangSA2022} with a responsivity of approximately $R_h \sim$ 0.20 mV. By dividing the photovoltage spectral density by the measured responsivity $R_h$, we are able to obtain the spectral density of the microwave electric field, thus yielding the bandwidth-normalized sensitivity spectrum.

To conduct a comprehensive noise analysis, we measure  the sensitivity spectra under a variety of experimental conditions, as depicted in Fig. 2d. The sensitivity spectrum corresponding to photodetector noise is flat within the frequency range of 10 kHz to 300 kHz, resulting in a sensitivity limit of $\mathrm{{NEF}_{pd}}$ = 3.0 nV cm$^{-1}$ Hz$^{-1/2}$. As displayed in the top figure, the sensitivity spectrum measured in the absence of atoms, attributable to the optical readout noise, is approximately three times greater than that associated with photodetector noise. This aligns with the estimated value of $\mathrm{{NEF}_{ph}}$ (see Methods). Most notably, the sensitivity spectrum measured in the presence of atoms, reflective of our setup's total noise, only exhibits a modest 15\% increase compared to measurements taken in the absence of atoms. This outcome markedly deviates from heterodyne detection based on atomic vapor cells, where random perturbations of the laser field owing to Rydberg dephasing or transit of thermal atoms typically cause sensitivity degradation\cite{JingNP2020} of over 10 dB within the frequency range of 10 kHz to 100 kHz. Presently, the sensitivity of our Rydberg receiver is predominantly limited by the probe laser shot noise. When the signal MW is turned on, the signal peak along with higher-order harmonic signals begin to emerge from a similar noise base as expected.

\begin{figure}[tb]
\begin{center}
\includegraphics[width=7.5cm]{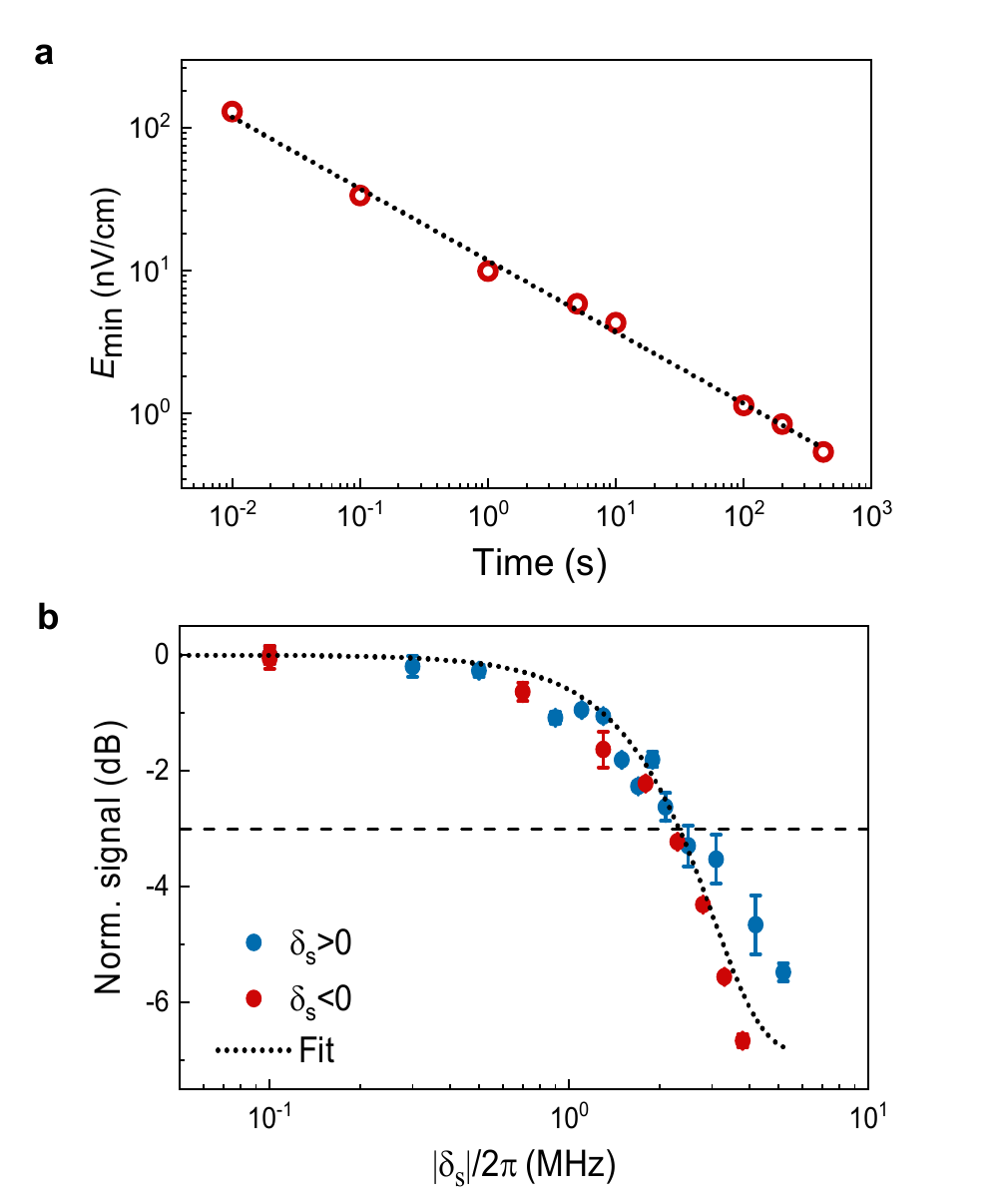}
\caption{ \textbf{Minimum detectable field and instantaneous bandwidth.}
\textbf{a}, Dependence of the minimum detectable field $E_{\min}$ on the total integration time $T'$. The temporal waveform of the noise base is averaged across a sequence of time traces. Following this, $E_{\min}$ is derived through dividing the DFT photovoltage by $\rm{R_h}$. This result agrees with the theoretical prediction (dotted line) of slope detection, scaling as $E_{\min }=\mathcal{S} / \sqrt{T'}$.
\textbf{b}, Normalized signal amplitude as a function of the MW detuning $\delta_{\rm{s}}$. The dotted curve signifies the Lorentz fit, indicating a 3-dB reduction in response at 2.3 MHz. }
\end{center}
\end{figure}

\bigskip
\noindent\textbf{Near-SQL electric field sensitivity.}
Subsequently, we adjust the incident power of the probe beam extensively to optimize the sensor's performance. As per the heating effect previously discussed, more potent incident light results in a more significant loss of atoms. As depicted in the inset of Fig. 3a, the atom number decreases to less than a tenth of the maximum when the probe power increases  from 0.33 $\mu$W to 9.53 $\mu$W. A pronounced inverse correlation exists between the shot noise linked with atoms and the probe laser. According to Eq. (1) and Eq. (10), we ascertain $\mathrm{{NEF}_{at}}$ and $\mathrm{{NEF}_{ph}}$ values, respectively, revealing a correlation coefficient of $r$ = -0.78 for the two noise equivalent fields. Fig. 3a presents the dependency of the measured sensitivity on the probe power. We observe that the field sensitivity initially improves as the probe power increases. However, a sensitivity decrease occurs around the 7.6 $\mu$W probe power, due to atom loss and Rydberg dephasing, which elevates $\mathrm{{NEF}_{at}}$ at higher probe powers.

Consequently, we set the incident probe power to 7.6 $\mu$W, which corresponds to a peak Rabi frequency of $\Omega^{(0)}_{\mathrm{P}}/2\pi$ = 4.7 MHz, and modulate the atom number by altering the gradient of the MOT quadrupole magnetic field. Fig. 3b demonstrates that enhancing the atom number improves  the electrometer sensitivity. The measurement aligns with the theoretical simulation of overall-noise limited sensitivity, exhibiting an asymptotic trend towards the sensitivity limited by thermal background noise. The atom number of $N$ = 5.2 $\times 10^{5}$ participating in the measurement leads to an optimal field sensitivity, achieving $\mathcal{S}$ = 10.0 nV cm$^{-1}$ Hz$^{-1/2}$. This is just 2.6 times greater than the SQL of $\mathrm{{NEF}_{at}}$ = 3.8 nV cm$^{-1}$ Hz$^{-1/2}$, and only 1.7 times greater than the room-temperature thermal noise limit of a lossless dipole antenna-based receiver\cite{IEEEreview2021}. To date, the measured sensitivity demonstrates the highest performance reported for atom-based MW sensing, signifying a threefold improvement over the previous heterodyne receiver\cite{PrajapatiAPL2021} based on a free-space coupled vapor cell. Additionally, Fig. 4a demonstrates the robust long-term stability of our system, attaining a minimum detectable MW field $E_{\min}$ = 540 pV cm$^{-1}$ for a measurement time of $T'$ = 420 s. Our cold atom receiver has a 2.3 MHz instantaneous bandwidth, comparable with that of a vapor cell-based receiver\cite{MeyerPRApplied2021} (Fig. 4b).

\bigskip
\noindent\textbf{Discussion}\\
Contrarily to the hot-atom method, our heterodyne receiver implements sequentially prepared cold atoms, substantially mitigating the transit noise and Doppler shift issues that degrade sensitivity. Along with the use of cold atoms, we set a moderate probe laser power which enhances the dominant photon shot noise in our system, consequently leading to a decrease in the SQL sensitivity due to atom loss and interaction-enhanced dephasing. Nevertheless, the receiver sensitivity is optimized by striking a balance between photon and atom shot noises. Theoretically, allowing more laser-cooled atoms to participate in detection at a higher probe power can elevate the receiver sensitivity. When compared to quantum sensors based on single\cite{FaconNature2016} or few particles\cite{NVSA2021}, our receiver, containing an ensemble of atoms, exhibits superior field sensitivity. The sensitivity of the receiver could be further enhanced by replacing the optical setup with a star-type three-photon excitation configuration\cite{ThreePH2016}, which significantly mitigates photon heating and retains more atoms in the MOT.  Furthermore, the three-photon scheme narrows the spectral linewidth and enhances the heterodyne responsivity. With these enhancements, the cold atom receiver's sensitivity is projected to rise, potentially exceeding the Johnson-Nyquist thermal noise limit of dipole antenna-based receivers.

Contrary to an atomic electrometer without local MW, where the unavoidable absorption loss in ladder-EIT establishes a distinct boundary for optimal sensitivity\cite{ShafferNP2012, MeyerPRA2021}, our approach offers considerable improvements. For instance, in a non-Doppler-broadened medium, the PSN limited sensitivity is at least 5.8 times\cite{MeyerPRA2021} inferior to the SQL sensitivity at an OD of 0.35. Differing from the three-level Rydberg-EIT scheme, our research leverages the atomic heterodyne technique to amplify the responsivity $R_{\rm{h}}$ and improve sensitivity. The PSN sensitivity in our system is merely a factor of roughly 2.4 greater than the SQL sensitivity, thus surpassing the PSN sensitivity limit inherent in the three-level ladder scheme. Additionally, the numerical model shows preliminarily promising prospects, demonstrating that ensuing enhancements could allow the cold-atom receiver to approach SQL at higher probe power levels and optimized optical depths (refer to Supplementary Section 2).

In summary, we have successfully mitigated the issues of spectrum Doppler-broadening and atomic transit noise in the Rydberg receiver, achieving exceptional sensitivity.
Recently, portable MOTs implementing the pyramidal\cite{ArltOC1998} or planar-integrated\cite{ChenPRA2022} configurations have been shown successful in capturing the atomic cloud at a comparable optical depth, heralding possible miniaturization of the cold atom receiver. Broadly, beyond detection of discrete frequencies, the cold Rydberg receiver introduced here is applicable to various heterodyne protocols for wideband electric-field sensing, such as through the ac Stark shift\cite{MeyerPRApplied2021} or two-photon resonance\cite{LiuPRA2022} in multilevel Rydberg atoms. Vetted with near-SQL sensitivity, this cold Rydberg receiver divulges potential applicability across disciplines, including radio astronomy\cite{RMNature2016, WangNA2019}, metrology\cite{StockMT2018, DingNP2022}, radar technologies\cite{DouglasNature2000}, atmospheric studies\cite{NjokuIEEE1982}, and even explorations of dark matter axions\cite{AxionPRL2017, JiangNP2021}.

\bigskip
\noindent\textbf{Methods}\\
{
\noindent\textbf{Optical readout noise.}
The noise sources within the optical readout path encompass both photodetector noise and photon shot noise. The photodetector noise establishes a sensitivity limit, denoted as ${\rm{NEF}}_{\rm{pd}}$. This can be represented in terms of the noise voltage of the photodetector ($\rm{NEV}$) and the responsivity of the heterodyne detection ($R_{\rm{h}}$) by the following equation:
\begin{eqnarray} \begin{array}{ll} {\rm{NEF}}_{\rm{pd}} = {\rm{NEV}} / {\rm{R_h}}. \end{array} \end{eqnarray}

The sensitivity ${\rm{NEF}_{\rm{ph}}}$ is determined by calculating the photon-shot-noise $P_N$ of the transmitted beam incident to the photodetector. It is calculated as:
\begin{eqnarray}\begin{array}{ll}P_N=\sqrt{P_0T_P/(\hbar \omega_P)} \hbar \omega_P=\sqrt{P_0T_P \hbar \omega_P},\end{array}
\end{eqnarray}
where $P_0$ represents the probe beam power prior to interaction with atoms, $T_P$ denotes the transmission through atomic medium, and $\omega_P$ is the angular frequency of the probe laser. Under the adiabatic approximation, a gradual change in signal MW amplitude, or $\Omega_S$, results in an alteration of the transmitted probe power, expressed as:
\begin{eqnarray}\begin{array}{ll}\Delta P = {P_0}{{\partial {T_P}} \over {\partial {\Omega _S}}}\Delta {\Omega _S}. \end{array}
\end{eqnarray}
The minimum detectable change in signal Rabi frequency $\Delta \Omega_S$ can be derived by equating the noise power with the alteration in transmitted power. This results in:
\begin{eqnarray}\begin{array}{ll}\Delta {\Omega _{S,\min }} = \sqrt {{{{T_P}\hbar \omega_P} \over {{P_0}}}} /{{\partial {T_P}} \over {\partial {\Omega _S}}}.\end{array}
\end{eqnarray}
Consequently, the PSN sensitivity can be expressed as:
\begin{eqnarray}\begin{array}{ll}{\rm{NEF}}_{\rm{ph}} =  {2\sqrt{2}\pi\hbar \over {{\mu_{\mathrm{MW}}}}} \sqrt {{{{T_P}\hbar \omega_P} \over {{P_0}}}}/{{\partial {T_P}} \over {\partial {\Omega _S}}}.
\end{array}
\end{eqnarray}
In this formula, the PSN-limited sensitivity improves as the incident probe power increases. We obtained $\mathrm{{NEF}_{ph}}$ = 9.1 nV cm$^{-1}$ Hz$^{-1/2}$ at an incident probe power of 7.6 $\mu$W, which aligns with the result of noise detection illustrated in Fig. 2d.

\bigskip
\noindent\textbf{ Noise correlation.}
Given a group of measurement results with respect to the $\mathrm{{NEF}_{ph}}$ and  $\mathrm{{NEF}_{at}}$, the correlation coefficient $r$ is defined as
\begin{eqnarray}
\begin{array}{ll}
r=\frac{\sum(\mathrm{{NEF}}_\mathrm{{at}}^{i}-\overline{\mathrm{{NEF}}}_\mathrm{{at}})(\mathrm{{NEF}}_\mathrm{{ph}}^{i}-\overline{\mathrm{{NEF}}}_\mathrm{{ph}})}{\sqrt{\sum(\mathrm{{NEF}}_\mathrm{{at}}^{i}-\overline{\mathrm{{NEF}}}_\mathrm{{at}})^2} \sqrt{\sum(\mathrm{{NEF}}_\mathrm{{ph}}^{i}-\overline{\mathrm{{NEF}}}_\mathrm{{ph}})^2}},
\end{array}
\end{eqnarray}
where $\overline{\mathrm{{NEF}}}_\mathrm{{at}}$ ($\overline{\mathrm{{NEF}}}_\mathrm{{ph}}$) represents the average of $\mathrm{{NEF}_{at}}$ ($\mathrm{{NEF}_{ph}}$) measurement.

\bigskip
\noindent\textbf{Effective gain of atomic sensor.}
Owing to the directional selectivity of atomic heterodyne detection, we can calculate  the effective reception gain $G$ of a free-space coupled Rydberg sensor for isotropic microwave fields:
\begin{eqnarray}
\begin{array}{ll}G = 4\pi /\int_0^{2\pi } {\int_0^\pi  {{F^{\rm{2}}}\left( {\theta ,\phi } \right)} } \sin \theta d\theta d\phi.
\end{array}
\end{eqnarray}
Here, the radiation pattern $F\left( {\theta ,\phi } \right)$, which receives the power from an incident plane wave with an angle of arrival $\left( {\theta ,\phi } \right)$, is given by:
\begin{eqnarray}
\begin{array}{ll}F\left( {\theta ,\phi } \right){\rm{ = }}\sin \theta  \rm{sinc}\left[ {{\pi L \over {{\lambda}}}\left( {\sin \theta \sin \phi  - \cos \beta } \right)} \right],
\end{array}
\end{eqnarray}
where $L$ represents the longitudinal length of the medium, $\lambda$ corresponds to the microwave wavelength, and $\beta$ is the angle between the propagation directions of the local microwave field and the laser beam. For the experimental parameters $L$ = 2 cm, $\lambda$ = 0.81 cm, and $\beta$ = 0, we obtain a gain of $G$ = 11.5.According to equation (\ref{eq:BBR}), the atomic sensitivity limited by thermal background at room temperature results in
$\mathrm{{NEF}_{ex}}$ = 3.2 nV cm$^{-1}$ Hz$^{-1/2}$.

\bigskip
\noindent\textbf{Atomic coherence time.}
The coherence time of the cold Rydberg ensemble used as the heterodyne sensor is defined by the formula:
\begin{eqnarray}
\begin{array}{ll}T_2=1 /\left(\gamma_0 + \gamma_t + \gamma_r + \Gamma_r/2 \right),
\end{array}
\end{eqnarray}
where $\gamma_0 \approx 2\pi \times$ 100 kHz is the single-atom Rydberg-EIT dephasing rate which arises from finite laser linewidth, blackbody radiation, and atomic interactions. The dephasing rate $\gamma_t$ refers to the rate caused by the transit of thermal atoms,  $\gamma_r$ is the inherent Rydberg dephasing associated with the dispersion of dipole-dipole interactions\cite{HanPra2016, BoulierPra2017, PetrosyanPRL2014} (see Supplementary Section 3), and $\Gamma_r$ indicates a negligible decay rate pertaining to the long-lived Rydberg state. It should be noted that we derived an effective dephasing rate by fitting the AT-splitting spectrum shown in Fig. 2b, which aligns well with the $T_2$ value calculated by Eq. (14).

\bigskip
\noindent\textbf{Thermal-noise limit of traditional receiver.}
For a lossless half-wave dipole antenna-based receiver, the electric-field sensitivity limited by Johnson-Nyquist thermal noise is defined as:
\begin{eqnarray}
\begin{array}{ll}
\mathcal{S}_\mathrm{th} =\sqrt{2 P_\mathrm{th} /\left(\epsilon_0 c A_d\right)},
\end{array}
\end{eqnarray}
where $P_\mathrm{th}=k_B T_{\mathrm{eq}}$ is the receiver noise power at an equivalent noise temperature $T_{\mathrm{eq}}$, $k_B$ is the Boltzmann's constant, and $A_d=0.41 \lambda^2 / \pi$ is the effective area of dipole antenna.

\bigskip
\noindent\textbf{Data availability}

The main data supporting the results in this study are available within the paper and its Supplementary Information. The raw datasets are available from the corresponding authors upon reasonable request.

\bigskip
\noindent\textbf{Code availability}

The codes used for the theoretical simulations are available from the corresponding authors upon reasonable request.

\bigskip
\bigskip\noindent\textbf{Acknowledgements}\\
\noindent We thank Weibin Li, Linjie Zhang, and Wenhui Li for fruitful discussions. The work is supported by the National Key Research and Development Program of China (Grants No. 2021YFA1402004 and 2022YFA1405300), the Key Area Research and Development Program of Guangdong province (Grants No. 2019B030330001 and No. 2020B0301030008); the National Natural Science Foundation of China (Grants No. 12225405, and No. U20A2074); Innovation Program for Quantum Science and Technology (Grant No. 2021ZD0301700), and the Guangdong Basic and Applied Basic Research Foundation (Grants No. 2021A1515010272, and No. 2023A1515030003).

\bigskip\noindent\textbf{Author Contributions}\\
\noindent H.T.T.,K.Y.L., and H.Y.  designed the experiment.
H.T.T., G.D.H., and S.Y.Q. carried out the experiments.
H.T.T., K.Y.L., Y.F.Z., and H.J. conducted raw data analysis.
K.Y.L., H.Y., and S.L.Z. wrote the paper, and all authors discussed the paper contents.
K.Y.L., H.Y. and S.L.Z. supervised the project.

\bigskip
\noindent\textbf{Competing Financial Interests}\\
\noindent The authors declare no competing financial interests.

\newpage

\section*{\large Supplemental Materials : Approaching the standard quantum limit of a Rydberg-atom microwave electrometer}


\renewcommand{\theequation}{S\arabic{equation}}
\setcounter{equation}{0}
\renewcommand{\thefigure}{S\arabic{figure}}
\setcounter{figure}{0}
\maketitle

\section*{SI. Theoretical model of Rydberg receivers }

In this section, we show how to derive the theoretical sensitivity for a Rydberg receiver implemented with the atomic gas at physical temperature $T_{\rm{a}}$. In atomic heterodyne scheme, we mainly focus on the microwave-dressed EIT spectra in the presence of interactions between Rydberg atoms\cite{JingNP2020}. Because of the low atomic density of sensing sample (i.e., far less than one atom within the blockaded sphere), we neglect the photon blockade effect, and phenomenologically quantify the spectral properties via the effective level shifts and dephasing rate in the theoretical model.

The time evolution of reduced density operator $\varrho$ obeys the optical Bloch equation
\begin{equation}
\partial_{t} \varrho=-\frac{i}{\hbar}\left[H, \varrho\right]+\mathcal{L}\varrho.
\end{equation}
For a four-level ladder-EIT diagram shown in Fig. 1b of the main text, the effective Hamiltonian $H$, based on the bare states \{${\left|1 \right\rangle}={\left|5S_{1/2} , F=2\right\rangle},{\left|2 \right\rangle}={\left|5P_{3/2}, F=3\right\rangle},{\left|3 \right\rangle}={\left|39D_{5/2} \right\rangle},{\left|4 \right\rangle}={\left|40P_{3/2} \right\rangle}$\}, is written as
\begin{equation}
\label{H}
\begin{aligned}
H=&-\hbar \sum_{k=2}^{4}\Delta_{k} A_{kk}-\frac{\hbar}{2}\left[\Omega_{{P}}  A_{21}+\Omega_{{C}}  A_{32}\right.\\
& +(\Omega_{L}+\Omega_{S} e^{-i \phi(t)})  A_{43}+\mathrm{H.c.}],
\end{aligned}
\end{equation}
where $ A_{ij}= |i\rangle\langle j|$ is the atomic transition operator, $\phi(t)=\delta_{s} t+\varphi_0$ is the time-dependent relative phase between the signal and local microwaves. In Eq. (S2), the detuning $\Delta_{k}(v)$ for atoms at a velocity $v$  is defined as
\begin{subequations}
\begin{align}
&\Delta_2(v)= \Delta_{\mathrm{P}}+ k_P v,\\
&\Delta_3(v)= \Delta_{\mathrm{P}} + V_{\mathrm{vdW}} + V_{\mathrm{dd}}+ k_P v-k_C v, \\
&\Delta_4(v)= \Delta_{\mathrm{P}} + V_{\mathrm{dd}} + V^{\prime}_{\mathrm{vdW}}+k_P v-k_C v+ k_L v,
\end{align}
\end{subequations}
where $ \Delta_{\mathrm{P}}$ is the probe laser detuning, $V_{\mathrm{vdW}}$ and $V^{\prime}_{\mathrm{vdW}}$ are the corresponding level shifts induced from van der Waals (vdW) interaction, and $V_{\mathrm{dd}}$ is the dipole-dipole exchange (DDE) shift, $k_{X}$ denotes the wave number of field $\Omega_{{X}}$ ($X \in \{{P}, {C}, {L} \}$). Here we ignore the Doppler shift of microwave field since it is much smaller than that of optical fields.

The Lindbladian term in Eq. (S1) describes the repopulation and relaxation processes of atomic system, which yields
\begin{equation}
\begin{aligned}
\mathcal{L} \varrho=&-\frac{\Gamma}{2}\left(A_{22} \varrho+\varrho A_{22}-2 A_{12} \varrho A_{12}^{\dagger}\right) \\
&-\frac{\Gamma_{r}}{2}\left(A_{33} \varrho+\varrho A_{33}-2 A_{23} \varrho A_{23}^{\dagger}\right) \\
&-\frac{\Gamma_{r^{\prime}}}{2}\left(A_{44} \varrho+\varrho A_{44}-2 A_{14} \varrho A_{14}^{\dagger}\right)\\
&-\sum_{k=3}^{4}(\gamma_0 + \gamma_t + \gamma_r)\left(A_{kk} \varrho+\varrho A_{kk}-2 A_{kk} \varrho A_{kk}\right),
\end{aligned}
\end{equation}
where $\Gamma$ is the decay rate of excited state $|2\rangle$, $\Gamma_{r}$, $\Gamma_{r^{\prime}}$ are the decay rates of long-lived Rydberg states $|3\rangle$, and $|4\rangle$ respectively, $\gamma_0 \approx 2\pi \times$ 100 kHz is the dephasing rate for single-atom Rydberg EIT,
\begin{equation}
\begin{aligned}
\gamma_t=\frac{1}{w \sqrt{2 \ln (2)}}\sqrt{\frac{8 k_B T_{\rm{a}}}{\pi m}} \nonumber
\end{aligned}
\end{equation}
denotes the atomic transit dephasing associated with beam waist $w$ and atom mass $m$, and $\gamma_r$ is the Rydberg dephasing rate owing to the dispersion of many-body interactions.

\begin{figure*}[tp]
\begin{center}
\includegraphics[width=17cm]{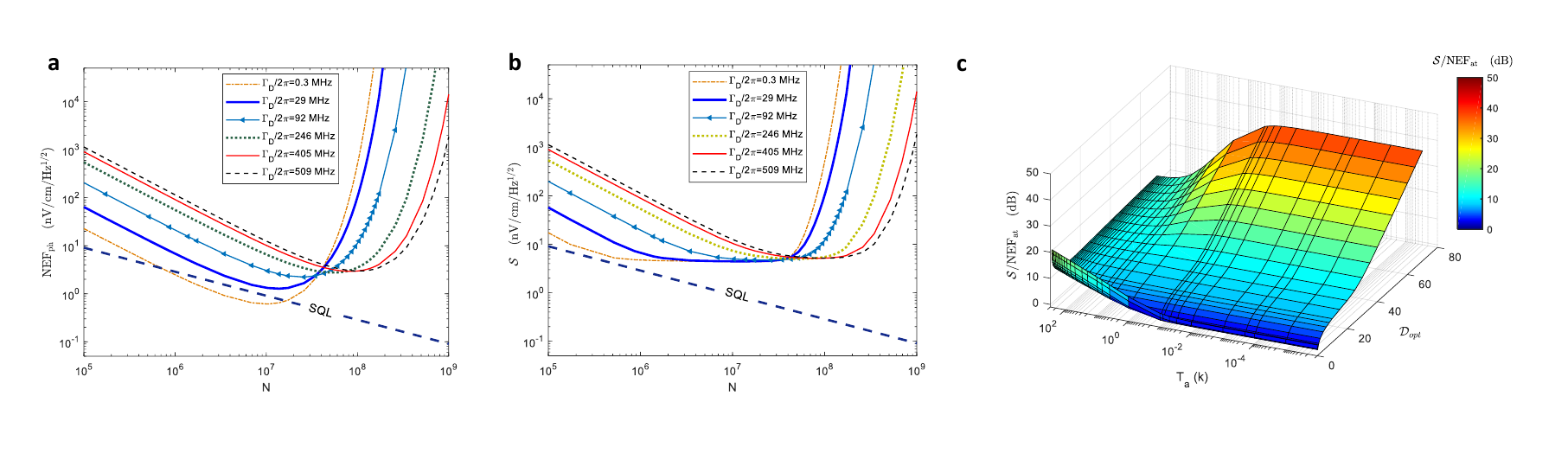}
\caption{\label{fig:compare} \textbf{Theoretical analysis of the sensitivity of Rydberg-atom microwave receiver.}
\textbf{a}, PSN limited sensitivity $\mathrm{NEF}_{\mathrm{ph}}$ as a function of the number $N$ of atoms in sensing volume.  \textbf{b}, Overall-noise limited sensitivity $\mathcal{S}$ versus total atom number $N$. $\Gamma_D$ is the Dopplor width of probe transition. The blue dashed line is the SQL sensitivity $\mathrm{NEF}_{\mathrm{at}}$ calculated by the equation (1) of the main text. \textbf{c}, The ratio of $\mathcal{S}$ to $\mathrm{NEF}_{\mathrm{at}}$ as a function of atomic temperature $T_{\rm{a}}$ and optical depth $D_{opt}$. The relevant parameters used in theoretical simulation is set as the same as the optimized parameters in experiment. }
\end{center}
\end{figure*}

To characterize the spectral response to microwave field, the Doppler-averaged four-level atomic coherence $\overline{\rho}_{21}$ derived from Eq. (S1) gives the probe transmission
\begin{equation}
\label{trans}
T_{P}={\mathrm{exp}}\{- \frac{D_{opt}\Gamma}{\Omega_P}  {\rm{Im}}[\overline{\rho}_{21}]\}.
\end{equation}
In Eq. (\ref{trans}), $D_{opt} = n_{at} \sigma_{12} L$ is the optical depth of atomic ensemble associated with the atomic density $n_{at}$ and the longitudinal length $L$, and
\begin{equation}
\sigma_{12} =\frac{2 k_{P} \left|\mu_{12}\right|^2}{\hbar \varepsilon_0 \Gamma}\nonumber
\end{equation}
is the absorption cross section on the transition $|1\rangle\leftrightarrow|2\rangle$, where $\mu_{12}$ is the electric dipole moment of probe transition. Considering the Doppler effect\cite{MeyerPRA2021}, the averaged matrix element $\overline{\rho}_{21}$ of atomic ensemble at the temperature $T_{\rm{a}}$ is given by
\begin{equation}
\overline{\rho}_{21}=\int_{-\infty}^{\infty} \frac{{\rm{exp}}^{(-v^2/u^{2})}}{u \sqrt{\pi}} \rho_{21}\left[\Delta_2(v), \Delta_3(v), \Delta_4(v)\right] dv,
\end{equation}
where $u =\sqrt{2{{k}_{B}}T_{\rm{a}}/{m}}$ is the most probable speed. To obtain a numerical solution, we truncate both the lower and upper limits of integral at the five-fold of the most probable speed.

In the adiabatic limit of Eq. (S1) when $|\Omega_{S}|, |\delta_{s}|\ll|\Omega_{L}|$, the spectral responsivity of probe laser on the signal microwave is a close approximation to that on the local microwave. Under the steady-state condition, we get the numerical solution to the optical Bloch equation even when the probe Rabi frequency is comparable with the coupling Rabi frequency. Based on this method, the photon-shot-noise (PSN) limited sensitivity is solved by a partial derivation at the optimal local microwave\cite{CompNoiseT2022}, and it writes
\begin{equation}
{\rm{NEF}}_{\rm{ph}} =  {2\sqrt{2}\pi\hbar \over {{\mu_{\mathrm{MW}}}}} \sqrt {{{{T_P}\hbar \omega_P} \over {{P_0}}}}/{{\partial {T_P}} \over {\partial {\Omega _L}}}  ,
\end{equation}
where $\omega_P$ and $P_0$ are the probe laser angular frequency and incident power respectively. Ultimately, we find the attainable sensitivity for heterodyne receiver
\begin{small}
\begin{equation}
\mathcal{S}=\sqrt{\mathrm{NEF}_{\mathrm{at}}^2+2r\mathrm{NEF}_{\mathrm{at}}\mathrm{NEF}_{\mathrm{ph}}+
\mathrm{NEF}_{\mathrm{ph}}^2+\mathrm{NEF}_{\mathrm{pd}}^2+\mathrm{NEF}_{\mathrm{ex}}^2}.
\end{equation}
\end{small}
As in the above form, the theoretical sensitivity at the optimal parameters in Fig. 3b of the main text is
\begin{widetext}
\begin{equation}
\begin{aligned}
\mathcal{S}&=\sqrt{3.70^2-2 \times 0.78 \times 3.70 \times 9.10 +9.10^2 +3.00^2+3.18^2}\; \mathrm{nV cm^{-1} Hz^{-1/2}}\\
&=7.94 \;\mathrm{nV cm^{-1} Hz^{-1/2}},
\end{aligned}
\end{equation}
\end{widetext}
and this theoretical result agrees with the measured sensitivity. Note that in this model we actually haven't considered the probe fluctuation caused by the transit of thermal atoms, which can degrade the theoretical sensitivity especially as the atomic temperature $T_{\rm{a}}$ increases.

\section*{SII. Comparison with the SQL sensitivity }

\begin{figure*}[tp]
\begin{center}
\includegraphics[width=18cm]{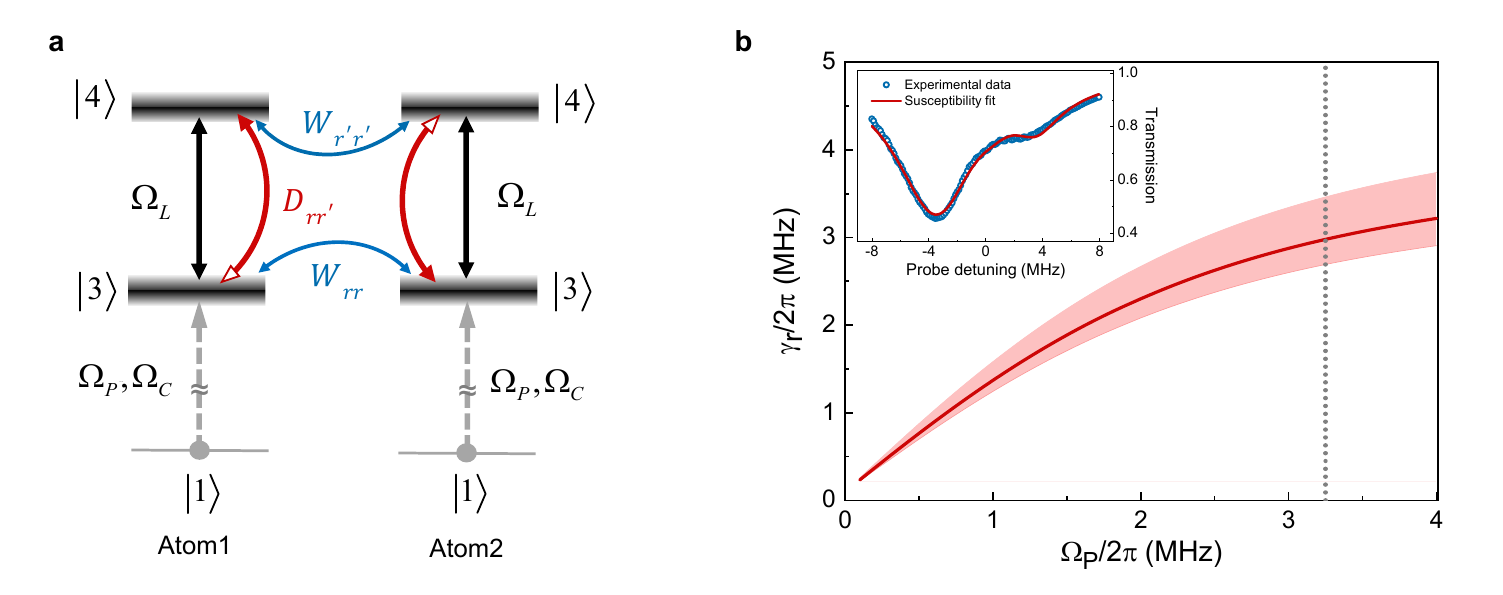}
\caption{\label{fig: interaction} \textbf{Energy-level shift and depahsing in an interacting Rydberg ensemble.}
\textbf{a}, Atoms 1 and 2 in the sensing volume interact via the $W_{rr}$ and $W_{r^{\prime}r^{\prime}}$ vdW interactions and the DDE interaction $D_{r r^{\prime}}$. In the microwave-dressed ladder-EIT system, the ground state $\left| 1 \right\rangle $ is coupled with Rabi frequencies ($\Omega_{\mathrm{P}}$  and $\Omega_{\mathrm{C}}$) to the Rydberg state $\left| 3 \right\rangle $ while a microwave field drives the Rydberg transition $|3\rangle\leftrightarrow|4\rangle$ with Rabi frequency $\Omega_{\mathrm{L}}$. The relative positions between energy levels $\left| 3 \right\rangle $ and $\left| 4 \right\rangle $ are only indicative. \textbf{b}, Interaction-induced dephasing rate $\gamma_r$ versus the probe Rabi frequency $\Omega_P$. The solid curve is the theoretical result from Eq. (S20), and the shaded zone represents the error band generated by the 5$\%$ uncertainty of optical depth and the anisotropy of $C_6$ interaction of the pairstate $\left|39D_{5/2} \right\rangle$. The probe Rabi frequency of our experiment is indicated with the vertical dotted line. The inset depicts transmission spectrum measured at the optimal point of atomic heterodyne, 
and the red curve is the fit of experimental data to the approximate susceptibility of Eq. (S21). }
\end{center}
\end{figure*}

In the following, we specifically discuss the dependance of the attainable sensitivity on the optical depth and physical temperature of the atomic sample. To better illustrate the influence of Doppler averaging, the atomic temperature is characterized as the Doppler width of the probe transition, the full width of which at half maximum is given by $\Gamma_D=2\sqrt{\ln 2}  k_P u $. Fig. S1(a) shows the PSN limited sensitivity as a function of the total atom number with $\Gamma_D/2 \pi$ = 0.3, 29, 92, 246, 405, and 509 MHz, respectively. As the Doppler width (or atomic temperature $T_{\rm{a}}$) increases from 0.3 MHz (100 $\mu$K) to 509 MHz (300 K), the effect of Doppler averaging on the optical depth induces a significant reduction in the PSN sensitivity, and the required optical depthes to provide the optimal sensitivities are increased. Moreover, for an optical thin ensemble of $D_{opt} <$ 10 (i.e. $N < 10^{7}$ in the sensing volume), the partial derivation $\partial T_P/\partial \Omega_L$ in Eq. (S7) is increased with the increase of optical depth, resulting in the improvement of the PSN sensitivity\cite{MeyerPRA2021} and even beating the SQL in a certain parameter region of the cold atom case [e.g. $\Gamma_D=2 \pi \times$ 0.3 MHz in Fig. S1(a)]. When the optical depth further increases, we can see that the PSN sensitivity, however, generally worsens due to the extreme absorption loss in the probe field.

In Fig. S1(b), we show the overall-noise limited sensitivity $\mathcal{S}$ obtained from Eq. (S8). In the presence of constant external noise, the low-temperature atomic sensitivities [such as the theoretical curves of $\Gamma_D/2 \pi$ = 0.3 MHz, and 29 MHz as shown in Fig. S1(b)] feature a flat bottom owing to the negative correlation between the two comparable equivalent fields $\mathrm{NEF}_{\mathrm{at}}$ and $\mathrm{NEF}_{\mathrm{ph}}$. The SQL result (i.e., $\mathrm{NEF}_{\mathrm{at}}$, blue dashed line) is determined assuming that all atoms in the sensing volume, where the atomic sample and all the control fields overlap, participate equally and the coherence time is $T_2=1 /\left(\gamma_0 + \gamma_t + \gamma_r + \Gamma_r/2 \right)$. Fig. S1(c) shows the ratio of $\mathcal{S}$ to SQL sensitivity as a function of the atomic temperature and the optical depth, illustrating that an optical thin ensemble of cold atoms is the only viable means of approaching the SQL in Rydberg electrometry.

\section*{SIII. Rydberg dephasing induced from atom-atom interactions }

In the above model of Rydberg electrometry, we have characterized the spectral properties with the additional energy shift and dephasing items compared to the single-atom model. Now we theoretically quantify the energy-level shift and dephasing due to Rydberg interactions. In the microwave-dressed Rydberg EIT, atoms in the states $\left|3 \right\rangle$ and $\left|4 \right\rangle$ interact with each other via the vdW potentials $W_{rr}=\hbar\left(C_6/ R^6\right) \hat{A}_{33}^1 \otimes \hat{A}_{33}^2$ and $W_{r^{\prime}r^{\prime}}=\hbar\left(C_6^{\prime} / R^6\right) \hat{A}_{44}^1 \otimes \hat{A}_{44}^2$, where $R$ is the interatomic separation. As illustrated in Fig. S2(a), the resonant Rydberg transition driven by microwave field gives rise to a dipole-dipole exchange interaction\cite{PetrosyanPRL2014} between the atoms $D_{r r^{\prime}}=\hbar\left(C_3/ R^3\right) \hat{A}_{34}^1 \otimes \hat{A}_{43}^2$.

For the $39D_{5/2}$ pairstate, the $W_{rr}$ vdW interaction is attractive and anisotropic, with the $C_{\rm{6}}$ coefficient for a strength range of 0.14-0.38 GH $\mu$m$^6$. Since the optical thin sample with a size of 4 $\times$ 4 $\times$ 20 mm$^{3}$ has a low atomic density $n_{at}$ = 5.68 (0.28) $\times$ {10$^{7}$}cm$^{ - 3}$, there is only $N_B=n_{at}\times 4 \pi R_B^3 / 3\approx 0.01$ atom within the blockade sphere, with a radius of $R_B=\left(2 {\overline{C}_{\rm{6}}} / \gamma_{\mathrm{EIT}}\right)^{1 / 6}$, where ${\overline{C}_{\rm{6}}}=\int_{0}^{\pi}C_6(\theta)\sin(\theta)d\theta$ is the averaged vdW coefficient, and $\gamma_{\rm{EIT}}\approx 2\pi\times$ 3 MHz is the EIT linewidth in our experiment. The $W_{r^{\prime}r^{\prime}}$ vdW interaction is unable to block the atomic excitation as well, and thus the ``blockade volume" in our model don't need to be treated as the superatom. Although the effect of photon blockade is negligible, there is still Rydberg dephasing left inherent to the dispersion of the energy level shifts and exchange coupling\cite{HanPra2016}.

In the following, we calculate the Rydberg dephasing $\gamma_r$ in the mean-field approach. The total vdW shift $\hat{V}_{\mathrm{vdW}}(\mathbf r)$ at the position $\mathbf r$ involves integration over the entire volume excluding the blockade sphere $V_B^{\mathbf r}=4 \pi R_B^3 / 3$. To approximately incorporate the spatial correlations between Rydberg polaritons\cite{WeimerPRL2008, DeSalvoPRA2016}, we introduce a short-range cutoff to the spatial integral at $R_B$, and the averaged energy level shift due to $W_{rr}$ interaction is
\begin{equation}
\begin{aligned}
V_{\mathrm{vdW}}&=  \overline{\sum_j \frac{C_6}{R_{i j}^6}} \approx \int_{R_B}^{\infty}\int_{0}^{\pi}\Sigma_{rr}n_{a t} \frac{C_6(\theta)}{r^6} 2 \pi r^2 \sin(\theta)d\theta d r \\
&= {{2\pi {\overline{C}_{\rm{6}}}\Sigma_{rr}{n_{at}}} \over {3R_B^3}},
\end{aligned}
\end{equation}
where the excitation fraction $\Sigma_{rr}$ within the blockade volume approximates the atom population at the $\left|3 \right\rangle$ level when excluding the many-body interactions\cite{PetrosyanPRA20013, BoulierPra2017},
\begin{small}
\begin{equation}
\Sigma_{rr}=\frac{\left(\Omega_{P} \Omega_{C} \Gamma_{r^{\prime}}\right)^{2}}{{\left(\Gamma \Gamma^{2}_{r}+\Omega_{L}^{2} \Gamma+\Omega_{C}^{2} \Gamma_{r^{\prime}}\right)}^{2}+\left(\Omega_{P} \Omega_{C} \Gamma_{r^{\prime}}\right)^{2}+\left(\Omega_{L} \Omega_{P} \Omega_{C}\right)^{2}}.
\end{equation}
\end{small}
The bandwidth of $W_{rr}$ vdW shift is defined as $2\sqrt{\vartheta _{rr}}$, where
\begin{equation}
\begin{aligned}
\vartheta _{rr}&=  \overline{(\sum_j \frac{C_6}{R_{i j}^6})^{2}} -V_{\mathrm{vdW}}^{2}\\
&\approx \int_{R_B}^{\infty}\int_{0}^{\pi} \Sigma_{rr} n_{a t} \frac{C_6^{2}(\theta)}{r^{12}} 2 \pi r^2 \sin(\theta)d\theta d r\\
&= {{2\pi{\Sigma_{rr}}{n_{at}}} \over {9R_B^9}}\int_{0}^{\pi}C^{2}_6(\theta)\sin(\theta)d\theta
\end{aligned}
\end{equation}
is variance of $\left|3 \right\rangle$ level shift from the mean value $V_{\mathrm{vdW}}$. In this way, we calculate the averaged energy level shift owing to the $W_{r^{\prime}r^{\prime}}$ interaction, which writes
\begin{equation}
V^{\prime}_{\mathrm{vdW}} \approx {{2\pi \Sigma_{r^{\prime}r^{\prime}}{n_{at}}} \over {3R^{\prime 3}_B}}\int_{0}^{\pi}{C^{\prime}_{\rm{6}}}(\theta)\sin(\theta)d\theta ,
\end{equation}
and the variance of the $\left|4 \right\rangle$ level shift is given by
\begin{equation}
\vartheta _{r^{\prime}r^{\prime}}\approx  \frac{{2\pi \Sigma_{r^{\prime}r^{\prime}}}{n_{at}}}  {9R_B^{\prime9}}\int_{0}^{\pi}C^{\prime2}_6(\theta)\sin(\theta)d\theta,
\end{equation}
where
\begin{small}
\begin{equation}
\Sigma_{r^{\prime}r^{\prime}}=\frac{\left(\Omega_{L} \Omega_{P} \Omega_{C}\right)^{2}}{{\left( \Gamma \Gamma^2_{r}+\Omega_{L}^{2} \Gamma+\Omega_{C}^{2} \Gamma_{r^{\prime}}\right)}^{2}+\left(\Omega_{P} \Omega_{C} \Gamma_{r^{\prime}}\right)^{2}+\left(\Omega_{L} \Omega_{P} \Omega_{C}\right)^{2}}
\end{equation}
\end{small}
is the excitation fraction of $\left|4 \right\rangle$ level, and
\begin{equation}
R_B^{\prime}=\left(2 \int_{0}^{\pi}{C^{\prime}_{\rm{6}}}(\theta)\sin(\theta)d\theta / \gamma_{\mathrm{EIT}}\right)^{1 / 6}
\end{equation}
is the corresponding blockade radius.

The resonant DDE interaction is relatively isotropic, and the exchange coupling with an eigenvalue $E_{rr^{\prime}}$ lifts the Rydberg energy levels, giving a DDE shift since $R_B\ll R_B^{\prime}$,
\begin{equation}
\begin{aligned}
V_{\mathrm{dd}}&=E_{rr^{\prime}}=  \overline{\sum_j \frac{C_3}{R_{i j}^3
}} \approx n_{a t}|\Sigma_{rr^{\prime}}|\int_{R_B}^{w} \frac{C_3}{r^3} 4 \pi r^2 d r\\
&= 4\pi {C_3}|\Sigma_{rr^{\prime}}|{n_{at}}\ln ({w \over {{R_B}}}),
\end{aligned}
\end{equation}
where $w$ is the $1/e^{2}$ radius of coupling laser beam, and
\begin{small}
\begin{equation}
|\Sigma_{rr^{\prime}}|=\frac{\Gamma_{r} \Omega_{P}^{2} \Omega_{C}^{2} \Omega_{L}}{{\left(\Gamma \Gamma^2_{r} +\Omega_{L}^{2} \Gamma+\Omega_{C}^{2} \Gamma_{r^{\prime}}\right)}^{2}+\left(\Omega_{P} \Omega_{C} \Gamma_{r^{\prime}}\right)^{2}+\left(\Omega_{L} \Omega_{P} \Omega_{C}\right)^{2}}
\end{equation}
\end{small}
is the probability amplitude of Rydberg transition. To determine the dephasing inherent to the dispersion of exchange coupling, we calculate the variance of resonant dipole-dipole interaction from the average value $E_{rr^{\prime}}$,
\begin{equation}
\begin{aligned}
\vartheta _{rr^{\prime}}&=  \overline{(\sum_j \frac{C_3}{R_{i j}^3})^{2}} -E_{rr^{\prime}}^{2}\\
&\approx \int_{R_B}^{w} |\Sigma_{rr^{\prime}}| n_{a t} \frac{C_3^{2}}{r^{6}} 4 \pi r^2  d r\\
&\approx {{4\pi C_3^{2}{|\Sigma_{rr^{\prime}}|}{n_{at}}} \over {3R_B^3}}.
\end{aligned}
\end{equation}

Therefore, the Rydberg dephasing inherent to the dispersion of many-body interactions is
\begin{equation}
\begin{aligned}
{\gamma _r} =  \sqrt {\vartheta _{rr}}  +\sqrt {\vartheta _{r^{\prime}r^{\prime}}}  + 2 \sqrt {\vartheta _{rr^{\prime}}}.
\end{aligned}
\end{equation}
Fig. S2(b) shows the dephasing rate as a function of the probe Rabi frequency. We find $\gamma _r$ = $2\pi\times$3.0 MHz at the optimal probe power. To confirm the mean-field calculation of Rydberg dephasing, we fit the measured transmission spectrum to an effective four-level susceptibility \cite{LiaoPRA2020} $\chi_{\rm{eff}}$,
\begin{equation}
\chi_{\rm{eff}}=  -\frac{2n_{0} \left|\mu_{12}\right|^2}{\hbar \varepsilon_0 } \frac{d_3 d_4-\Omega_{\mathrm{L}}^2 / 4}{d_2 d_3 d_4-d_2 \Omega_{\mathrm{L}}^2 / 4-d_4 \Omega_C^2 / 4},
\end{equation}
where $d_j=\Delta_{j}-i \gamma_{j}$ is the complex detunings with $\gamma_{j}$ being the total dephasing rate of state $\left|j \right\rangle$. The inset shows the fitted profile only with the dephasing rates $\gamma_{3}$ and $\gamma_{4}$ as free parameters. We find that the theoretical result of Rydberg dephasing is in good agreement with the dephasing extracted from the susceptibility fit.



\end{document}